\documentclass[sigconf, natbib=true]{acmart}

\acmSubmissionID{XXX}

\usepackage[skip=0pt]{caption}
\usepackage{balance}
\usepackage{tabularx}

\usepackage[inline]{enumitem}

\AtBeginDocument{%
  \providecommand\BibTeX{{%
    \normalfont B\kern-0.5em{\scshape i\kern-0.25em b}\kern-0.8em\TeX}}}

\setcopyright{acmcopyright}
\copyrightyear{2023}
\acmYear{2023}
\setcopyright{rightsretained}
\acmConference[SIGIR '23]{Proceedings of the 46th International ACM SIGIR Conference on Research and Development in Information Retrieval}{July 23--27, 2023}{Taipei, Taiwan} \acmBooktitle{Proceedings of the 46th International ACM SIGIR Conference on Research and Development in Information Retrieval (SIGIR '23), July 23--27, 2023, Taipei, Taiwan}\acmDOI{10.1145/3539618.3591923}
\acmISBN{978-1-4503-9408-6/23/07}

\begin{document}

\title{\workshopName{} @ SIGIR 2023: The First Workshop on Generative Information Retrieval}

\author{Gabriel Bénédict}
\email{g.benedict@uva.nl}
\orcid{0000-0002-3596-0285}
\affiliation{%
  \institution{University of Amsterdam and RTL NL}
  \country{The Netherlands}
}

\author{Ruqing Zhang}
\email{zhangruqing@ict.ac.cn}
\orcid{0000-0003-4294-2541}
\affiliation{ 
    \institution{ICT, Chinese Academy of Sciences} 
    \country{China}
}

\author{Donald Metzler}
\email{metzler@google.com}
\affiliation{ 
    \institution{Google Research} 
    \country{USA}
}

\renewcommand{\shortauthors}{Gabriel Bénédict, Ruqing Zhang, \& Donald Metzler}

\newcommand{\citetemp}{\textcolor{red}{[cite]}}
\newcommand{\gabriel}[1]{\textcolor{magenta}{[GAB: #1]}}
\newcommand{\ruqing}[1]{\textcolor{blue}{[RUQING: #1]}}
\newcommand{\workshopName}{Gen-IR}



\begin{abstract}
Generative information retrieval (IR) has experienced substantial growth across multiple research communities (e.g., information retrieval, computer vision, natural language processing, and machine learning), and has been highly visible in the popular press. Theoretical, empirical, and actual user-facing products have been released that retrieve documents (via generation) or directly generate answers given an input request. We would like to investigate whether end-to-end generative models are just another trend or, as some claim, a paradigm change for IR. This necessitates new metrics, theoretical grounding, evaluation methods, task definitions, models, user interfaces, etc. The goal of this workshop\footnote{https://coda.io/@sigir/gen-ir} is to focus on previously explored Generative IR techniques like document retrieval and direct Grounded Answer Generation, while also offering a venue for the discussion and exploration of how Generative IR can be applied to new domains like recommendation systems, summarization, etc. The format of the workshop is interactive, including roundtable and keynote sessions and tends to avoid the one-sided dialogue of a mini-conference.

\end{abstract}


\maketitle

\section{Title}

\workshopName{} @ SIGIR 2023: The First Workshop on Generative Information Retrieval

\section{Motivation}

Last year saw the rise of generative IR on two fronts. We will refer to them as:
\begin{enumerate*}[label=(\roman*)]
    \item~\emph{Generative Document Retrieval (GDR)}: via a generative process, retrieve a ranked list of existing documents (e.g. Wikipedia or news articles) that match a query and
    \item~\emph{Grounded Answer Generation (GAG)}: retrieve a human readable generated answer that matches a query; the answer can link to or refer to a document.
\end{enumerate*}

On the GDR end of the spectrum, \citeauthor{rethinkingSearch} first proposed an end-to-end \emph{model-based retrieval} approach in a position paper~\cite{rethinkingSearch}: directly predict identifiers of candidate documents, instead of indexing all documents (a.k.a. \emph{index-retrieve-then-rank}). The position paper builds on generative entity linking~\cite{autoregressiveRetrieval, memorizationGeneration, BERTKNN}, later extended for long sequences~\cite{longSequences}. The generative model is expected to embed all relevant information that is in the documents. Soon after, \citeauthor{differentiableIndex} released Differentiable Search Indexes (DSI), the first model generating indexes of Wikipedia articles~\cite{differentiableIndex}. The above mentioned position paper~\cite{rethinkingSearch} goes beyond GDR, towards GAG and full-fledged end-to-end retrieval models that generate answers.
    
On the GAG end of the spectrum~\cite{hybridRetrievalConv, chatbot}, recent Large Language Models (LLMs) have been released to the public that are essentially (conversational) IR models. Some are conversational with aspects of reinforcement learning (ChatGPT\footnote{\url{https://openai.com/blog/chatgpt/}} or Claude\footnote{\url{https://www.anthropic.com/constitutional.pdf}}),
some cite their sources (Phind\footnote{\url{https://phind.com/about}} or Perplexity\footnote{\url{https://www.perplexity.ai/}}), some are focused on science (Galactica\footnote{\url{https://galactica.org/}}), some can do all of the above and more (YOU\footnote{\url{https://you.com/}}), and others have yet to be released (Sparrow\footnote{\url{https://www.deepmind.com/blog/building-safer-dialogue-agents}}).

Generative IR as an end-to-end model has clear benefits over the \emph{index-retrieve-then-rank} paradigm.
\begin{enumerate*}[label=(\roman*)]
    \item It is simpler and more flexible.
    \item The training pipeline is compressed.
    \item There is no need for an index of documents that is tedious to query or compute similarity with.
\end{enumerate*}
But Generative IR also comes with its challenges. Namely,
\begin{enumerate*}[label=(\roman*)]
    \item it has yet to be demonstrated that retrieval performance is improved on big datasets (such as the full MS-MARCO dataset~\cite{MSMARCO}),
    \item generative models can hallucinate (i.e., generate false information). This is more obviously true for LLMs that generate answers (GAG) than for retrieval models that generate doc-ids (GDR).
    \item The \emph{infinite index} paradigm~\cite{infiniteIndex}: if LLMs can generate an infinite amount of answers to a given query, then classic recall-based IR evaluation metrics like NDCG cannot rely on a finite amount of true positives. 
\end{enumerate*}

A workshop on Generative IR will question whether IR is truly facing a paradigm change at the theoretical level~\cite{rethinkingSearch}. This event will also be a way to reflect on Generative IR's benefits and challenges, as retrieval-like LLMs (GAG) get released to the general public. Finally, we will encourage submissions and discussions on further Generative IR topics and models, where existing literature is scarce, such as recommender systems, Learning to Rank, diffusion models, etc. We compiled a list of related literature\footnote{\url{https://github.com/gabriben/awesome-generative-information-retrieval}}. %

\section{Theme and purpose of the workshop}

\workshopName{} 2023 will be a forum for discussion about the challenges in applying (pre-trained) generation models for information retrieval as well as the theory behind the models and applications. 
The aim of this workshop is multi-fold: 
\begin{enumerate*}[label=(\roman*)]
    \item discussing the main challenges in designing and applying generative retrieval models in practice,
    \item establishing a bridge for communication between academic researchers and industrial researchers around Generative IR, 
    \item providing an opportunity for researchers to present new directions and early insights, and
    \item creating an agenda for Generative IR according to the 4 pillars bellow (Model Architecture, Training, Evaluation, Applications). This agenda will then ideally be periodically revised at future occurrences of the workshop. 
\end{enumerate*}

Our call for papers and the theme of the panel / roundtable discussions will evolve around these 4 pillars. For now Generative IR revolves mostly around Generative Document Retrieval (GDR) and Grounded Answer Generation (GAG). We leave space for further tasks in the 4th pillar.

\subsection{Model Architecture} Despite the preliminary studies on pre-trained language models (PTMs) for GDR, most research in this direction focuses on straightforwardly applying existing PTMs that are specifically designed for NLP into IR applications such as T5 \cite{raffel2020exploring} and BART \cite{lewis2019bart}.  
These encoder-decoder architectures do not consider the IR cues that might benefit the downstream IR tasks, such as GAG. These cues include information about ranking, entity disambiguation, and the causal relationships behind ranking tasks.

Another solution could be to generate documents via other types of models that can provide a range of predictions, like diffusion models~\cite{diffusion}. Diffusion models have already been tested for language generation and categorical data in general~\cite{diffusionCategorical} and are thus candidates for both GDR and GAG tasks.

\subsection{Training} 

Despite the strong experimental performance of GDR models, the potential of generative models for general search problems is limited by the training strategies that are currently employed.

\begin{itemize}[leftmargin=*]

\item \textbf{Learning To Rank objective.} Traditional \emph{index-retrieve-then-rank} paradigm implies a Learning To Rank objective at the end of the pipeline. This objective is commonly expressed as point-wise, pair-wise, or list-wise. Following the new \emph{model-based retrieval} paradigm, the objective is global over the whole corpus and usually defined as a standard seq2seq objective, i.e., maximizing the output doc-id likelihood with teacher forcing conditioned on the query. There are many interesting questions to help understand whether such optimization is optimal, how it connects with existing Learning to Rank paradigms, and so on.

\item \textbf{Generalization Ability.} So far most studies only demonstrate the effectiveness of their approaches on retrieval datasets where a query has only one relevant document. 
In the future, we should extend the generalization ability of GDR to different search tasks, including a query with a relevant document, with multiple relevant documents at one relevance grade, and with multiple relevant documents at different relevance grades. 
One option to predict multiple documents via  model based retrieval is to use contrastive learning between the document and query representations. 

\item \textbf{Incremental Learning.} For GDR models, there remain open questions about the practical applicability of such models to dynamic corpora. In dynamic and open IR system, documents are incrementally added or removed from the indexed corpus. It is valuable to explore continuously updated learning objectives over new or removed documents (e.g. \cite{DSI}). 

\end{itemize}

\subsection{Evaluation} We consider several topics for the evaluation of Generative Document Retrieval and Grounded Answer Generation:
\begin{itemize}[leftmargin=*]
    \item We are not aware of an evaluation on a big dataset for either GDR or GAG (such as the full MS-MARCO dataset~\cite{MSMARCO}).
    \item Evaluation metrics need to be designed taking into account the specifics of the generative paradigm. These metrics should ideally both suit traditional IR and Generative IR.
    \item Human evaluation of Generative IR is still at its infancy. Note that ChatGPT leverages Reinforcement Learning with Human Feedback (RLHF)~\cite{RLHF,stiennon2020learning}, while Claude uses RL from AI Feedback (RLAIF)~\cite{Claude}.
    \item Interpretability and causality are still hard to determine. In the context of GAG, this implies citing its sources, a.k.a. \emph{attribution}~\cite{attribution} (via for example a citation token~\cite{galactica}). In other words bridging the gap between GDR and GAG.
    \item Robustness to adversarial attacks (how easy is it to create fake facts or fool the Grounded Answer Generation model) and to distribution shifts (does transfer learning across datasets work?). 
    \item Efficiency of models. GDR requires considerably less compute power than GAG. Is there a way to bring computational costs down for GAG or to provide more information with the same amount of compute with GDR (e.g. a ranking of documents instead of just one document or a summary of documents)?
    \item GAGs tend to be very assertive about their claims. Uncertainty estimates would be particularly desirable for GAGs and especially for the ones which don't cite their sources like ChatGPT.
    \item GAGs can appear like they have a mind of their own. Some new conceptual metrics and learning constraints have been proposed like truthfullness, harmlessness, honesty and helpfulness~\cite{Claude}.
\end{itemize}

\subsection{Applications} At inference time, both GDR and GAG are sensitive to prompting strategies. Given particular prompts, it has been shown that one can provoke ChatGPT into hallucinating answers. As a solution, could we use a generation model to unify GDR and GAG, so as to provide document references to source material making it much easier to highlight the authoritativeness / accuracy of the answer? 

Furthermore, there are several applications to Generative IR that have not yet been subject to much scrutiny beyond GDR and GAG. We can think of summarization, Knowledge-Intensive Language Tasks (KILT) (e.g.~\cite{factVerification,chen2022corpusbrain}), recommender systems (e.g.~\cite{generativeSlate, generativeReco}) and learning to rank (e.g.~\cite{generativeRanking}).

\section{Format}

\workshopName{} will be an interactive full-day hybrid workshop that avoids the one-sided dialogue of a mini-conference.

\begin{itemize}[leftmargin=*]
    \item Invited panel (industrial and academic) [hybrid]. Candidates from different institutions and companies accepted our invitation: Neeva, Google, Meta AI, Tsinghua University, Chinese Academy of Science, Sapienza University of Rome, Samaya AI, KAIST, University of Waterloo, Huggingface, Stanford University.    
    \item Contributed paper presentations as posters [onsite] and video demos [online].
    \item An interactive session to share lessons learned [hybrid].
    \item Breakout sessions on issues that emerge from the contributed papers and demos (to be determined after the submission deadline but prior to the workshop) [onsite].
\end{itemize}

\subsection{Workshop schedule}

\subsubsection*{Morning}

\phantom{}

\noindent
\begin{tabular}{ll}
\toprule
\textbf{Time} & \textbf{Activity} \\ 
\midrule
08.30–08.45 & Opening \\ 
08:45–09:15 & Panel Discussions (academic) \\ 
09:15–10:00 & Poster Session - (1) Model Architecture \\ 
10:00–10:30 & Coffee break \\ 
10:30–11:00 & Panel Discussions (industrial)  \\ 
11:00–11:45 & Poster Session - (2) Training \\ 
11:45–12:15 & Breakout preparation \\ 
11:45–13:30 & Lunch \\ 
\bottomrule
\end{tabular}

\subsubsection*{Afternoon}

\phantom{}

\noindent
\begin{tabular}{ll}
\toprule
\textbf{Time} & \textbf{Activity} \\ 
\midrule
13:30–14:00 & Panel Discussions - Setting an agenda for Gen-IR \\ 
14:00–14:45 & Poster Session - (3) Evaluation \\ 
14:45–15:30 & Refreshment break \\ 
15:30–16:30 & Breakout \\ 
14:00–14:45 & Poster Session - (4) Applications \\ 
17:15–17:30 & Round up and closing discussions \\ 
\bottomrule
\end{tabular}

\subsubsection*{Schedule}

\phantom{}

\noindent
\begin{tabularx}{\columnwidth}{lX}
\toprule
\textbf{Date} & \textbf{Event} \\ 
\midrule
May 2, 2023 & Submission deadline \\ 
Jun 14, 2023 & Notification \\ 
Jul 1, 2023 & Camera ready versions of accepted papers due \\ 
Jul 27, 2023 & Gen-IR workshop \\ 
\bottomrule
\end{tabularx}

\section{Organizers}

\textbf{Gabriel Benedict} is an industry PhD candidate at University of Amsterdam, in collaboration with RTL NL. He is doing a mix of theoretical and applied AI research. The main themes are metrics-as-losses for neural networks, normative diversity metrics for news recommendation, intent-satisfaction modelling, video-to-music AI and most recently diffusion for IR tasks.

\textbf{Ruqing Zhang} is an associate professor in Institute of Computing Technology (ICT), Chinese Academy of Sciences (CAS). She has worked on a number of problems related to natural language generation and neural ranking models. Her current research is especially how to design generative models for IR, how to improve the robustness of ranking models, and how to make IR  trustworthy with the lens of ``causality''. 

\textbf{Donald Metzler} is a Senior Staff Research Scientist at Google Inc. Prior to that, he was a Research Assistant Professor at the University of Southern California (USC) and a Senior Research Scientist at Yahoo!. He currently leads a research group focused on a variety of problems at the intersection of machine learning, natural language processing, and information retrieval. He is a co-author of the position paper~\cite{rethinkingSearch}. 

\section{PC members}

Potential PC members for reviewing paper submissions: 

\begin{itemize}

    \item Andrew Yates,  University of Amsterdam
    \item Arian Askari, Leiden University
    \item Hainan Zhang, JD
    \item Hyunji Lee, KAST AI
    \item James Thorne, KAIST AI
    \item Nicola De Cao, University of Amsterdam
    \item Qingyao Ai,  Tsinghua University
    \item Roi Cohen, Tel Aviv University
    \item Ronak Pradeep, University of Waterloo
    \item Sheng-Chieh Lin, University of Waterloo
    \item Shengyao Zhuang, The University of Queensland
    \item Vinh Q. Tran,  Google Research
    \item Xiao Wang, University of Glasgow
    \item Xinyu Ma, Baidu
    \item Yujia Zhou, Renmin University of China
    \item Zhicheng Dou,  Renmin University of China

\end{itemize}

\section{Selection Process}

We will solicit submission of papers of two to six pages through an open call for papers, representing reports of original research, preliminary research results, proposals for new work, descriptions of generative models based toolkits tailored for IR, and position papers.

All papers will be peer reviewed by the program committee and judged by their relevance to the workshop, especially to the two main themes, and their potential to generate discussion.

\section{Target Audience}

The target audience is the broad range of researchers in industry and academia interested in IR and especially in Generative IR. We will advertise the workshop via a dedicated website and a Twitter/Mastodon account.
\section{Related Workshops}

As an emerging paradigm, there have not been related workshops held previously at SIGIR or other conferences.

\bibliographystyle{ACM-Reference-Format}
\balance
\bibliography{sample-base}

\end{document}